\documentclass[aps,prl,twocolumn,superscriptaddress,groupedaddress,balancelastpage]{revtex4}
\usepackage[compat=1.0.0]{tikz-feynman}
\usepackage{footmisc}
\usepackage{amssymb}
\usepackage{multirow}
\usepackage{bm}
\usepackage{amsmath}
\usepackage{graphicx}
\usepackage{epstopdf}
\usepackage{subfigure}
\usepackage{natbib}
\usepackage{epsfig}
\usepackage{amsfonts}
\usepackage{mathrsfs}
\usepackage[toc,page,title,titletoc,header]{appendix}
\usepackage[colorlinks,urlcolor=black, linkcolor=blue,citecolor=blue,anchorcolor=blue]{hyperref}
\usepackage{dsfont,amsthm,amsbsy}

\usepackage{fancyhdr}
\usepackage{braket}

\usetikzlibrary{quotes,angles}
\newcommand{\obtuseangle}{\kern.08em
\begin{tikzpicture}
    \draw coordinate (a) at (0.14,0);
    \draw coordinate (b) at (0,0);
    \draw coordinate (c) at (-.12,0.18);
    \draw (a) -- (b) -- (c) pic [draw=black]{} ;
\end{tikzpicture}
\kern.08em
}

\begin{document}
\title{Pair-density-wave and reentrant superconducting tendencies originating from valley polarization}

\author{Zhaoyu Han}
 \affiliation{Department of Physics, Stanford University, Stanford, California 94305, USA}

\author{Steven A. Kivelson}
  \affiliation{Department of Physics, Stanford University, Stanford, California 94305, USA}
\begin{abstract}
We investigate superconducting pairing tendencies of a two-dimensional electron fluid with both valley and spin degrees of freedom,  both  without and in the presence of an in-plane magnetic field. We present suggestive theoretical arguments that spontaneous valley polarization can lead to exotic singlet superconducting tendencies, including pair-density-wave order at zero field and re-entrant superconductivity at high field. We also find a reduced magnetic response in the polarized valley, which allows a finite violation of the Pauli limit. These results are obtained by a mean-field approach to a generalized $t$-$J$ model on a triangular lattice and in the dilute limit. Phenomenological similarities to results of recent magic-angle twisted trilayer graphene experiments
are noted.
\end{abstract}

\maketitle

Valley polarization refers to an imbalanced electron occupancy in two or more separated low-energy regions in the Brillioun zone (BZ). It can be achieved with various dynamical techniques in valleytronics~\cite{schaibley2016valleytronics,PhysRevMaterials.4.104005, liu2019valleytronics, PhysRevB.92.121403} for two-dimensional materials. In recent studies of layered and especially moir\'e flatband materials, spontaneous valley polarization has been invoked in many circumstances both theoretically~\cite{Dodaro_2018,PhysRevX.8.031089,PhysRevB.102.201107,PhysRevB.102.085103,lian2020tbg,PhysRevX.9.031021, kang2021cascades,PhysRevB.100.155426,PhysRevB.100.035448,parker2020strain,christos2021correlated}, and in experiments~\cite{zondiner2020cascade,saito2020isospin,PhysRevB.102.081403,yu2021correlated}. A recent study of the strong-coupling Holstein-Hubbard model on a triangular ladder has found spontaneous valley polarization when the effective bandwidth is sufficiently suppressed~\cite{huang2021pair} by poloranic effects. Valley polarization, induced by either intrinsic correlations or extrinsic environmental couplings, thus can be generically expected to be more significant in systems with reduced kinetic energy. Meanwhile, various superconducting (SC) phases are found to be present in such systems~\cite{ PhysRevLett.125.167001,  huang2021pair, cao2018unconventional,park2021tunable, cao2021pauli,hao2021electric,kim2021spectroscopic}. In this study, we investigate the implication of co-existing valley and spin polarization on SC pairing instabilities. 

In the context of BCS theory, the nature of the preferred SC state is determined by both an effective pairing interaction vertex, and a generalized (bare) pairing susceptibility of the normal state, $\chi^0$. Both of them can be thought as matrices, the indices of which represent the states of the cooper pairs, including their relative and center-of-mass momentum, $\vec{k}$ and $\vec{q}$. Assuming the relevant interaction that may lead to SC state is structureless attractions within energy shells of width $\Lambda_0$ around the Fermi surfaces (FS), the main determining factor is then $\chi^0$, which only depends on the (possibly strongly renormalized) band-structure and $\vec{q}$ of the cooper pairs. The largest $\chi^0(\vec{q})$ thus corresponds to the leading SC instability in the system, the divergence of which is usually associated with the nesting between two Fermi surfaces (FSs), i.e. $\epsilon(\hat{k}) =  \epsilon(\vec{q}-\hat{k})$ for $\hat{k}$ on one of the FSs. Once its value exceeds the inverse of the effective attraction strength, the pairing strength $|\Delta|$ at zero temperature will be determined by the attraction strength, the physical cutoff $\Lambda_0$, and the densities of states of the paired FSs.

In this letter, we present a mean-field analysis of a simple $t$-$J$-$V$ model on a triangular lattice in the dilute limit, in which both valley and spin degrees of freedom are active. We obtain a rich phase diagram at zero in-plane magnetic field $H$, where various orders compete as a function of $V$ and $J$. Particularly, we identify a broad valley polarized phase regime. Inside a subregion of this phase, we analyse the response behavior of $H$ and discuss the implication of the renormalized band structure on different pairing susceptibilities.  The main findings are summarized as follows: 1) At zero field $H=0$,  we find comparable inter- and intra-valley pairing susceptibilities that diverges in the dilute limit, with the result that the preferred singlet SC state {\em can be} intra-valley paired with a non-zero Cooper pair momentum, $\vec q=\pm 2 \vec{K}$ with $\vec{K}$ being the wavevector of K point in the Brillouin zone, i.e. pair-density-wave (PDW) phases~\cite{agterberg2020physics} can be favored.
2) In the presence of weak $H$, due to the delicate interplay among multiple ordering tendencies in the system, the effective Land\'e g-factor in the polarized valley is reduced. This results in a weaker magnetic response of the FSs and thus an enhanced upper critical field for intra-valley paired states at low-field, i.e. a violation of the nominal Pauli limit $H_c(T=0) \sim |\Delta(T=0,H=0)|/(g\mu_B)$~\cite{chandrasekhar1962note, PhysRevLett.9.266}. 3) At higher field, spin polarization order competes with the valley polarization before it completely wins. There inevitably exists a critical field strength at which a pair of FSs with both spin and valley indices different are re-balanced, resulting in a divergence of the singlet inter-valley, $\vec{q} = 0$ pairing susceptibility. This opens the possibility of re-entrant SC at high field. Both of observations 2) and 3) are reminiscent of observations in an experiment of magic-angle twisted trilayer graphene (MATTG)~\cite{cao2021pauli}.

{\it Model and method: } The Hamiltonian we study reads:
\begin{align}
    \mathcal{H} &= \hat{P}_G(\mathcal{H}_0 + \mathcal{H}_{\text{int}})\hat{P}_G \\
    \mathcal{H}_0 &= - t \sum_{\langle i,j\rangle, \sigma} \left(c^\dagger_{i\sigma} c_{j\sigma} +\text{h.c.}\right) -\mu \sum_i n_i \nonumber\\
    & \ \ \ \ \ \ \ \ \ \ \ \  - h \sum_i (n_{i\uparrow} - n_{i\downarrow}) \\
    \mathcal{H}_{\text{int}} &= \sum_{\langle i,j \rangle  }  J(\vec{S}_i\cdot\vec{S}_j- \frac{n_i n_j}{4})+V n_i n_j
\end{align}
where $c_{i\sigma}$ annihilate a spin-$\sigma$ electron on site $i$ ($\updownarrow$ represent align or anti-align with $\vec{H}$), $\vec{S}_i$ is the spin operator of site $i$, $\langle i,j\rangle$ represent the nearest neighbor pairs of sites and $h\equiv g \mu_B H/2$ represent the effect of an in-plane magnetic field. $\hat{P}_G$ is the projection operator that enforces a no-double-occupancy constraint. Although this $t$-$J$-$V$ model has a direct correspondence to a strong-coupling limit of the Holstein-Hubbard model~\cite{PhysRevLett.125.167001}, we regard it as a phenomenological model that can potentially be relevant to various strongly correlated systems. Thus, we consider a wide range of parameters that may correspond to different scenarios. For example, $V$ is the sum of multiple contributions, including the Coulomb repulsion and phonon-mediated attractions, so it can be positive or negative; similarly, $J$ can vary in a wide range due to the possible coexistence of anti-ferromagnetic super-exchange processes and the ferromagnetic exchange integral of Coulomb potential between neighboring Wannier orbits. Moreover, either due to the polaronic effects or moir\`e band flattening, we are motivated to consider the parameter regimes where the strengths of the interactions are comparable with that of the hopping element. Despite the presence of `infinite' on-site repulsion and the intermediate interactions, a density-matrix-renormalization-group (DMRG) study has found sharp Fermi points on ladders for this model~\cite{huang2021pair}. This encourages us to suggest that the 2D case may be similarly well described from a mean-field perspective -  at least there is no reason to expect it to be worse than the 1D case.

To realize a simple model with valley degrees of freedom, we consider a system on a triangular lattice with {\it negative} nearest-neighbor hopping elements $t<0$, in the dilute electron limit~\footnote{We also note that a particle-hole transformation $c_{i\sigma}^\dagger\rightarrow c_{i\sigma}$ can reverse the sign of $t$ and change the density $n_{i\sigma} \rightarrow  1 - n_{i\sigma}$, while leaving other terms invariant. Therefore, the dilute {\it electron} limit with a negative $t$ is equivalent to dilute {\it hole} limit with positive $t$.}. In this case, the electrons are expected to be confined in the neighborhoods of two band minima, i.e. valley centers, located at the $K$ and $K'$ points in the Brillioun zone. We then propose the ansatz state for the ground state of the system, which is specified by a set of variational parameters $\vec{n} = (n_{K\uparrow},n_{K\downarrow},n_{K'\uparrow},n_{K'\downarrow})$, i.e. the electron occupation densities with different combinations of valley and spin indices. We will refer to those combinations as flavors. This method renders the calculations more straightforward and the physical insights more transparent than the standard Hartree-Fock treatment presented in supplementary materials~\footnote{See Supplemental Material for the descriptions of the equivalent Hartree-Fock treatment, explicit expressions of the eigenvalues of the energy matrix, and the analysis of magnetic responses for the other subregions of the valley polarized phase.}, and we have checked that this method produces equivalent results. At $T=0$, the mean-field variational free energy that we wish to minimize can be expressed as an expansion in powers of the electron density:
\begin{align}
    E_{\text{var}} \equiv& \langle \vec{n}|\mathcal{H}|\vec{n}\rangle = \vec{\xi}^T \cdot \vec{n}  + \vec{n}^T \cdot M \cdot \vec{n} +\mathcal{O}(n^3) \\
    \vec{\xi} \equiv& (\mu+h,\mu-h,\mu+h,\mu-h)\\
    M\equiv& \frac{1}{2\rho_0} \tau^0 \otimes \sigma^0 + (3|t|+3V-3J)\tau^0\otimes\sigma^1\nonumber\\
    & \ \ + \frac{9}{2}V\tau^1\otimes\sigma^0 + (3|t|+3V-\frac{3}{4}J)\tau^1\otimes \sigma^1 
\end{align}
where we have defined the density of states per valley per spin $\rho_0\approx 0.1 / |t|$, redefined the chemical potential such that it is measured from the bare band bottom $-3|t|$, and adopted Pauli matrices $\sigma^{i=0,1}$ and $\tau^{i=0,1}$, acting on spin and valley index respectively, to simplify the expression. $\sigma^0$ and $\tau^0$ are simply identity matrices, and $\sigma^1$ or $\tau^1$ flips the spin or valley index. Since we are considering the dilute electron limit, i.e. the total electron density $n\rightarrow 0$ limit, we can discard the cubic and higher order terms in $n$, which simplifies the analysis greatly.  Note that the projection operators have introduced effective interactions of strength $\sim|t|$ into the system~\footnote{The effect of $\hat{P}_G$ can be accounted for by recognizing $\hat{P}_Gc^\dagger_{i\sigma}c_{j\sigma}\hat{P}_G=(1-n_{i\bar{\sigma}})c^\dagger_{i\sigma}c_{j\sigma}(1-n_{j\bar{\sigma}})$ and $\hat{P}_Gn_{i\sigma}\hat{P}_G=n_{i\sigma}(1-n_{i\bar{\sigma}})$.}. In the expression for the variational energy, it is now straightforward to identify the $\tau^0 \otimes \sigma^1$ or $\tau^1 \otimes \sigma^0$ terms as the interactions between electrons with spin or valley index flipped, and $\tau^1 \otimes \sigma^1$ as the interaction between electrons with both spin and valley indices flipped. Therefore, when those effective interactions are strong enough compared to the inverse of the density of states, the ``normal'' state, which has an equal density of each flavor, becomes unstable by a mechanism similar to Stoner ferromagnetism, and the system can spontaneously polarize into a symmetry broken state. To determine the ground state, we diagonalize the energy matrix $M$ as:
\begin{align}
    M &= \sum_{\alpha = 0, \text{S}, \text{V}, \text{SVL}}\lambda_\alpha \vec{\eta}^T_\alpha \vec{\eta}_\alpha
\end{align}
with eigenvectors
\begin{align}
    \vec{\eta}_0= (1,1,1,1)/2 &, \  \vec{\eta}_\text{S} = (1,-1,1,-1)/2 \nonumber\\
     \vec{\eta}_\text{V}= (1,1,-1,-1)/2 &, \   \vec{\eta}_\text{SVL} = (1,-1,-1,1)/2.
\end{align}
and eigenvalues that can be straightforwardly obtained (see supplementary materials for explicit expressions). Note that since the interaction terms are traceless, the sum of all four eigenvalues is simply $2/\rho_0$. Here, S, V and SVL respectively stands for spin, valley and spin-valley locked polarizations. It is natural to decompose the antsatz state $\vec{n}$ in this basis as $\vec{n} = \sum_\alpha m_\alpha \vec{\eta}_\alpha$, and interpret $m_{\alpha\neq 0}$ as the order parameter of the corresponding order. The variational energy can thus be reexpressed as:
\begin{align}
    E_\text{var} = & \lambda_0 \left(\frac{n}{2}-\frac{\mu}{\lambda_0}\right)^2 + \lambda_\text{S}\left (m_\text{S} - \frac{h}{\lambda_\text{S}}\right)^2 \nonumber\\
    & \ + \lambda_\text{V} m^2_\text{V} +\lambda_\text{SVL} m^2_\text{SVL}
\end{align}
Now the problem turns into an optimization problem with contraints that each component of $\vec{n}$ is non-negative, and all the components sum up to half the total density $n$, which can be fixed to a small number by tuning $\mu$. For any set of parameters, this is an easily solvable problem.

{\it Phase diagram at zero field. } At zero field, we note that any partially polarized state is unfavorable, since the energy monotonically depends on the absolute values of the order parameters $|m_{\alpha\neq 0}|$; the energy is minimized when  all $m_{\alpha\neq 0}$ either vanishes or take the maximally allowed value, $n/2$. Therefore, we can list all the candidate symmetry broken states as follows. Firstly, there can be S, V or SVL polarized states with only the corresponding order parameter $|m_{\alpha}|=n/2$, whose energies relative to the unpolarized normal state are $\lambda_{\alpha} n^2/4$. Secondly, all three order parameters can simultaneously have absolute value $n/2$ with sign structure such that $m_\text{S}m_\text{V}=m_\text{SVL} n/2$. The consequence is that all the electrons are polarized into a single combination of spin and valley, i.e. this is a flavor (F) polarized state with energy $\lambda_\text{F} n^2/4$ with $\lambda_\text{F}\equiv \lambda_\text{S}+\lambda_\text{V}+\lambda_\text{SVL}$. Whenever the smallest of these four energies is negative, the system will spontaneously polarize into the corresponding state.

In the above analysis, we have implicitly assumed that a uniform ground state is stable at the given density $n$. However, when $\lambda_0$ is negative, or $\lambda_0$ plus any one of the other three eigenvalues is negative, the system is no longer stable against spontaneously increasing total density. At fixed mean electron density, this implies phase separation (PS) between a high density region and a vacuum region. The dilute limit analysis in principle fails in such circumstances, but it is plausible to suppose that the `high' density region remains  moderately dilute, and hence forms the same symmetry broken state predicted by the previous analysis. 

\begin{figure}[t]
\includegraphics[width=\linewidth]{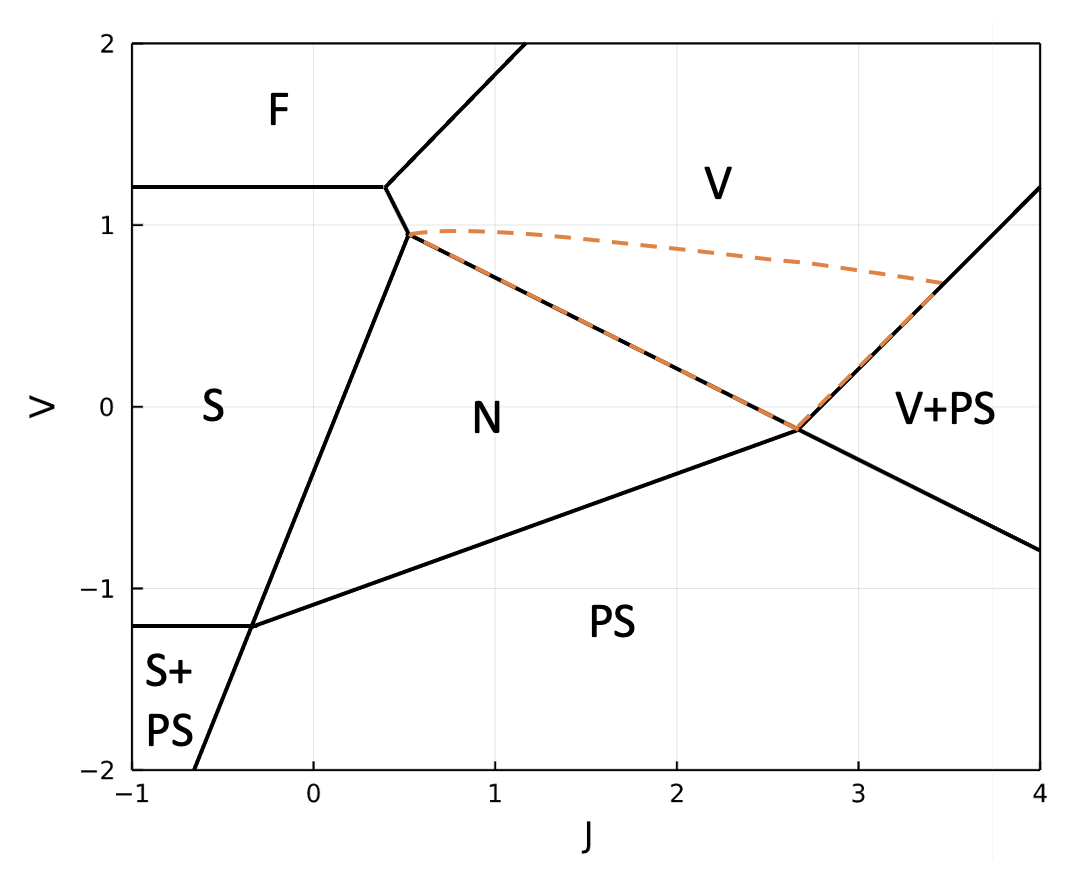}
\caption{The phase diagram at zero temperature and zero in-plane magnetic field. $V$ and $J$ are measured in units of $|t|$. F, S, V, N, PS stands for flavor polarized, spin polarized, valley polarized, normal and phase separated. S+PS or V+PS labels regions of phase separation in which the higher density phase is likely S or V, respectively. The orange dashed lines enclose the region where the magnetic response is discussed.}
\label{PD}
\end{figure}

We summarize the phase diagram at zero field derived from the above analysis in Fig.~\ref{PD}. All the phase transitions are first-order, although one should bear in mind that the approximation of discarding $\mathcal{O}(n^3)$ terms fails near those transition lines.  The inclusion of higher order terms could possibly introduce small partially polarized regions between the normal and fully polarized phases, and make the transitions continuous. 

{\it Pairing susceptibilities and magnetic response of valley polarized states. } At each point of the phase diagram, the renormalized band structure can be obtained, so the SC instability that would arise from additional weak attractive interactions can be explored. Since we are considering the dilute limit in which all the FSs are nearly circular, the most divergent SC susceptibilities correspond to the nesting between FSs. Generally, the leading pairing susceptibility between two flavors, $a$ and $b$, would have the BCS logarithmic divergence were their FSs identical upon inversion. So if they are of similar size, the diverging part of their contribution to the susceptibility can be approximated as:
\begin{align}
    \chi^0(\vec{q}_{ab}) \approx \rho_0 \ln\frac{\Lambda_0 \overline{|\epsilon^F|}}{\delta^2_{ab}}\ .
\end{align}
Here $\vec{q}_{ab}$ is the averaged momentum displacements of the pockets of flavors $a$ and $b$, $\Lambda_0$ is an appropriate UV cut-off determined by the physical nature of the attractive interactions, $\overline{|\epsilon^F|}$ is an average of absolute values of the Fermi energies, which are defined by the energy difference between the renormalized band bottom and the chemical potential, and $\delta_{ab}$ is the energy mismatch of the two FSs, which is defined by an average of $|\epsilon_a(\vec{q}_{ab}-\hat{k}_b)|$ with $\hat{k}_b$ on the FS of b (or vice versa). When any of the Fermi energies turn negative so that the corresponding FS is absent, $\chi^0$ simplifies to $\rho_0 \ln \Lambda_0/\delta_{ab}$. Within the approach we are adopting, these quantities for a state specified by density distribution $\vec{n}$ can be calculated as:
\begin{align}
    \epsilon_a^F &= n_a/\rho_0- \frac{\partial E_\text{var}}{\partial n_{a}}\bigg{|}_{\vec{n}}\\
    \delta_{ab} &= |\epsilon_a^F-\epsilon^F_b|
\end{align}
Whenever $\delta$ vanishes in this estimation, higher order effects in $n$, e.g. trigonal warping, should in principle be considered.

For the following discussion, we will consider the case of a valley polarized phase and an additional interaction that favors singlet pairing. Without loss of generality, we assume the electrons occupy the $K$ valley hereinafter. 

At zero field, the energy mismatch between $K\uparrow$ and $K\downarrow$ vanishes to the leading order, while the trigonal warping of the FSs yields $\delta_{\text{intra}}= c n^{3/2} /\rho_0$ with $c$ an $\mathcal{O}(1)$ constant. The resulting intra-valley susceptibility then can be calculated as $\chi^0(\vec{q}=2\vec{K})\approx 2\rho_0 \ln(\sqrt{\rho_0 \Lambda_0} / c n) $. Meanwhile, despite the fact that the states in valley $K'$ lie slightly above the chemical potential, the pair-fluctuations into and out of it are still allowed. The inter-valley pairing susceptibility corresponding to uniform SC is $\chi^0(\vec{q}=0)\approx 2\rho_0 \ln (\Lambda_0/\delta_{\text{inter}})$ with $\delta_{\text{inter}} = (|\lambda_\text{V}|+1/2\rho_0)n$ (the factor of $2$ in $\chi^0$ comes from two pairs of FSs for this $\vec{q}$). Therefore, both susceptibilities diverges with roughly the same asymptotic behavior as $n\rightarrow 0$, in contrast with the unpolarized case where the inter-valley nesting is always exact regardless of $n$. Which SC state is preferred thus depends on the detailed structure of the added attractions.

\begin{figure}[t]
\includegraphics[width=\linewidth]{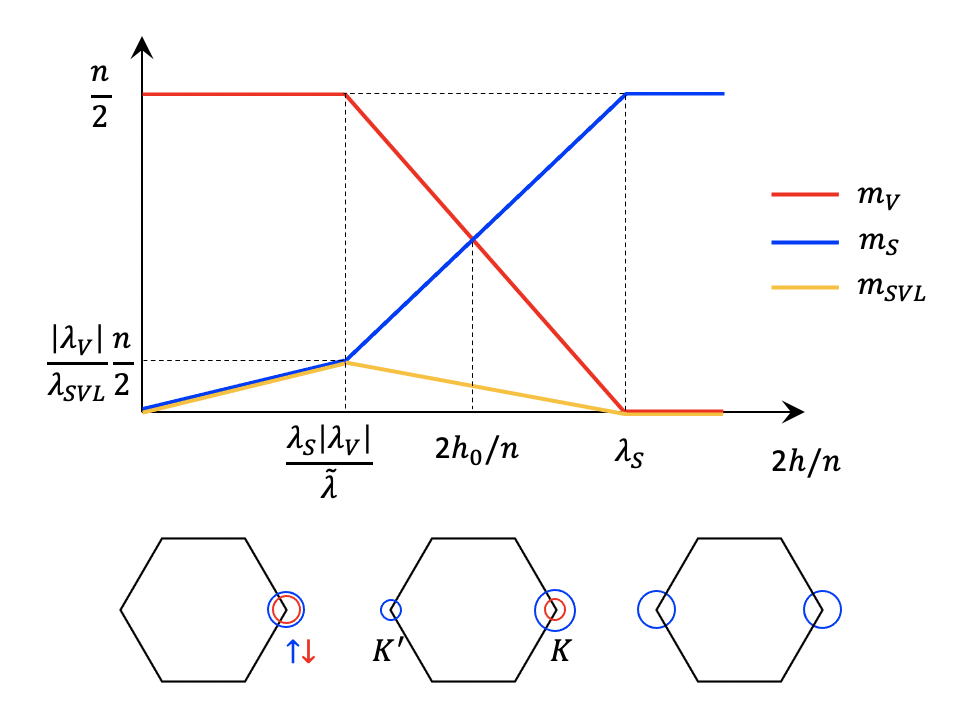}
\caption{Upper panel: the magnetic response of the order parameters in a subregion of the valley polarized phase, in which $\tilde{\lambda} \equiv \frac{\lambda_\text{S}\lambda_\text{SVL}}{\lambda_\text{S}+\lambda_\text{SVL}}>|\lambda_\text{V}|$. At $h=h_0$, $m_\text{S}=m_\text{V}$ implies $n_{K\uparrow}=n_{K'\downarrow}$. Lower panel: illustration of the FSs in the three stages of magnetization. The black hexagons represent the boundary of the first Brillioun zone.}
\label{MR}
\end{figure}

We now consider the effect of an in-plane magnetic field, $h\neq 0$, and analyze the resulting partially polarized states. Although the magnetic response at every point in the phase diagram can be analyzed straightforwardly, for the sake of simplicity and relevance to our topic, we will focus on a subregion of the V phase, corresponding to the region enclosed by the orange dashed line in Fig.\ref{PD}, in which $\tilde{\lambda} \equiv \frac{\lambda_\text{S}\lambda_\text{SVL}}{\lambda_\text{S}+\lambda_\text{SVL}}>|\lambda_\text{V}|$. Some qualitative discussions of the magnetic response in other subregions can be found in supplementary materials. In Fig.~\ref{MR}, we summarize the three-stage magnetic response  of the order parameters as a function of increasing $h$, as well as representative illustrations of the FSs in different ranges of $h$. In the first stage $h<\frac{n}{2}\frac{|\lambda_\text{V}|\lambda_\text{S}}{\tilde{\lambda}}$, the system is partially spin polarized, but remains fully valley polarized, until the magnetic field is large enough to start populating the spin-$\uparrow$ electrons in the $K'$ valley. In this stage, the mismatch between the two FSs in the $K$ valley can be calculated as
\begin{align}\label{weakfield}
    \delta_{\text{intra}} = \frac{g\mu_B H}{\rho_0 (\lambda_\text{S}+\lambda_\text{SVL})} \equiv \tilde{g}\mu_B H\ .
\end{align}
Comparing this with the case of non-interacting electrons, the Land\'e $g$-factor is effectively renormalized by a factor that is {\it less} than one in the entire region we are considering as long as $2\rho_0|\lambda_\text{V}|<1$ (which is true for the model studied), and reaches its minimum value, $g/2$, near the transition line between V and V+PS phases specified by the condition $\lambda_0+\lambda_\text{V} = 2/\rho_0 - (\lambda_\text{S}+\lambda_\text{SVL}) = 0$. The FSs are thus less sensitive to the in-plane magnetic field, allowing an $\mathcal{O}(1)$ violation of the Pauli limit if the zero-field state were an intra-valley paired SC. 

Upon further increasing the magnetic field  in the range $({n}/{2})\left[|\lambda_\text{V}|\lambda_\text{S}/ \tilde{\lambda}\right]\leq h \leq (n/2) \lambda_\text{S} $, $n_{K'\uparrow}$ rapidly increases from zero, and $n_{K\downarrow}$ gets depleted. Part way through this interval,  there must
occcur a critical magnetic field $h_0$ at which $n_{K'\uparrow}=n_{K\downarrow}$, so that the energy mismatch between the two FSs of those flavors vanishes as $\sim n^2$ and the inter-valley pairing susceptibility strongly diverges. This opens the possibility of SC pairing at a high field that is completely independent of the zero-field pairing strength $|\Delta(H=0)|$. Actually, as $h$ increases from $0$ to $h_0$, the inter-valley susceptibility grows, so if this pairing developed at zero field, this allows an arbitrarily large violation of Pauli limit. On the other hand, assuming intra-valley pairing was preferred at zero field and it has been eliminated at $h_0$, this implies re-entrant SC with a different pair momentum. Finally, for $h > (n/2) \lambda_S$, the spins are fully polarized and all singlet SC tendencies are suppressed.

{\it Discussions and outlook. } In this letter we have presented a concrete model that exhibits spontaneous valley polarization and a rich variety of SC tendencies. However, the method adopted and the qualitative results obtained in our letter can be readily generalized to other systems with valley degree of freedom, even for multi-valley systems on any two-dimensional lattice. Specifically, in an appropriate dilute limit, one can explore the effect of the interactions and identify the varieties of possible generalized ferromagnetic orders in the system using the method employed in the first part of this letter. For bi-valley systems, if there is a valley polarized phase, most of the results obtained in the second part can be directly applied, since there we made little reference to the detailed structure of the band and interactions for the specific model and lattice. This suggests a relatively robust route to PDW order \footnote{This mechanism is distinct from a intra-valley triplet pairing and a inter-valley singlet pairing mechanism proposed for a spin-valley locked system~\cite{hsu2017topological,Venderleyeaat4698}}, and may be relevant to some recent suggestive evidences of PDW order in vanadium-based Kagom\`e metals~\cite{chen2021roton}. We emphasize that all the basic ideas we discussed from a weak coupling perspective can be corroborated - at least on ladders - by DMRG results on a similar model~\cite{huang2021pair}, which do not rely on approximations. More generally, while Hartree-Fock approximation is surely not justified when the interactions are strong, it is well recognized as a sensible first step in identifying possible phases and behaviors.

Noticing that the valley polarization order and SVL order are rather similar in that they both do not respond directly to the magnetic field, another direct generalization of our results would be for the SVL phase. Although our current model turns out not to have a SVL phase region, it is easy to modify it to stabilize such a phase. In a SVL phase, the discussion would be totally the same as long as we interchange $\lambda_\text{V}$ and $\lambda_\text{SVL}$ as well as `inter' and `intra'. The only differences would be that, the inter-valley nesting becomes exact at zero-field, making intra-valley pairing unlikely in this case. 

We would like to point out the phenomenological similarities between our findings and a recent MATTG experiment~\cite{cao2021pauli}, where a large but {\it finite} violation of the Pauli limit of the zero-field SC state, and a re-entrant SC order at high field was observed at a relatively low filling. The phase transition between them, if there is a direct one, seems to be first order. All of these observations are consistent with the qualitative results obtained in this letter.

{\it Note added. } While we conduct this research, we became aware of some recent work~\cite{PhysRevLett.127.097001,PhysRevB.104.174505}, which explain the MATTG experiment from different perspectives by invoking triplet pairing. We were partially motivated by Ref.~\cite{PhysRevB.104.174505} to add the discussion about the SVL phase.

{\it Acknowledgement. } We thank Hong Yao for illuminating discussions.  SAK was supported, in part, by NSF grant No. DMR-2000987 at Stanford.

\bibliographystyle{apsrev4-1}
\bibliography{valley}

\section*{Supplementary Materials}

\section{Hartree-Fock treatment}

\label{HF}
Here we provide an equivalent Hartree-Fock treatment of the system, by which the complete renormalized band structure can be obtained. In this approach, we decompose the interaction as:
\begin{align}
    \mathcal{H}_{\text{int}} \rightarrow  
    &-\frac{J}{2}  \sum_{\langle i,j \rangle , \sigma} \left( \langle c^\dagger_{i\sigma} c_{j\sigma} \rangle c^\dagger_{j\Bar{\sigma}} c_{i\Bar{\sigma}} +  c^\dagger_{i\sigma} c_{j\sigma} \langle c^\dagger_{j\Bar{\sigma}} c_{i\Bar{\sigma}}\rangle \right)  \nonumber \\
    &-V\sum_{\langle i,j \rangle , \sigma} \left( \langle c^\dagger_{i\sigma} c_{j\sigma} \rangle c^\dagger_{j{\sigma}} c_{i{\sigma}} +  c^\dagger_{i\sigma} c_{j\sigma} \langle c^\dagger_{j{\sigma}} c_{i{\sigma}} \rangle\right) \nonumber\\ & +(V-\frac{J}{2})  \sum_{\langle i,j \rangle , \sigma} \left(\langle n_{i\sigma} \rangle n_{j\Bar{\sigma}} +  n_{i\sigma} \langle n_{j\Bar{\sigma}} \rangle \right)- E_\text{int}
\end{align}
where 
\begin{align}
    E_\text{int} =& \sum_{\langle i,j \rangle , \sigma} \big{[} (V-\frac{J}{2}) \langle n_{i\sigma} \rangle \langle n_{j\Bar{\sigma}}\rangle -  \frac{J}{2} \langle c^\dagger_{i\sigma} c_{j\sigma} \rangle \langle c^\dagger_{j\Bar{\sigma}} c_{i\Bar{\sigma}}\rangle\nonumber\\
   & \ \ \ \ \ \ \ \ - V\langle c^\dagger_{i\sigma} c_{j\sigma} \rangle\langle c^\dagger_{j{\sigma}} c_{i{\sigma}} \rangle \big{]}
\end{align}
The Gutzwiller projection can be neglected for the interaction terms since its effects on those terms are of order $\mathcal{O}(n^3)$ and thus can be neglected in dilute limit. Meanwhile, the effects of Gutzwiller projection are non-negligible for the Fermion bilinear terms:
\begin{align}
    \hat{P}_G \mathcal{H}_0\hat{P}_G &= -t\sum_{\langle i,j\rangle , \sigma} \left[(1-n_{i\bar{\sigma}})c^\dagger_{i\sigma}c_{j\sigma}(1-n_{j\bar{\sigma}}) +\text{h.c.} \right] \nonumber\\
   & \ \ \ \ \ \ \  -\mu \sum_{i} n_{i\sigma}(1-n_{i\bar{\sigma}}) - h\sum_i(n_{i\uparrow}-n_{i\downarrow}) \nonumber\\
    &\rightarrow  \mathcal{H}_0 + t\sum_{\langle i, j\rangle, \sigma} (n_{i\bar{\sigma}}+n_{j\bar{\sigma}})(\langle c^\dagger_{i\sigma}c_{j\sigma} \rangle + \text{c.c.}) \nonumber\\
    & \ \ \ \ \ \ \ + t\sum_{\langle i, j\rangle, \sigma} (\langle n_{i\bar{\sigma}}\rangle + \langle n_{j\bar{\sigma}}\rangle )( c^\dagger_{i\sigma}c_{j\sigma}  + \text{h.c.})\nonumber\\
    & \ \ \ \ \ \ \ + \mu \sum_{i} (n_{i\uparrow} \langle n_{i\downarrow}\rangle + \langle n_{i\uparrow}\rangle n_{i\downarrow}) - E_0
\end{align}
where
\begin{align}
    E_0 &= t\sum_{\langle i, j\rangle, \sigma} \langle n_{i\bar{\sigma}}+n_{j\bar{\sigma}}\rangle (\langle c^\dagger_{i\sigma}c_{j\sigma} \rangle + \text{c.c.}) \nonumber\\
    & \ \ \ \ \ + \mu \sum_i \langle n_{i\uparrow}\rangle \langle n_{i\downarrow}\rangle 
\end{align}
To focus on the physics of valley and spin polarization, we seek for uniform solutions that can break time-reversal, spin-rotation and inversion symmetries. Specifically, we look for the self-consistent solution of mean-field Hamiltonian:
\begin{align}\label{eq:MFH}
    H_\text{MF} &= \sum_{i,a,\sigma} \left( \tilde{t}_\sigma c^{\dagger}_{i,\sigma} c_{i+\hat{e}_a,\sigma} +\text{h.c.}\right) -
     \sum_{i,\sigma} \tilde{\mu}_\sigma n_{i\sigma}
\end{align}
with mean-field equations
\begin{align}\label{mfeqs}
\tilde{t}_\sigma &= -t - V \langle c^{\dagger}_{i+\hat{e}_1,\sigma} c_{i,\sigma}\rangle - \frac{J}{2} \langle c^{\dagger}_{i+\hat{e}_1,\bar{\sigma}} c_{i,\bar{\sigma}}\rangle + 2 t n_{\bar{\sigma}}  \\
\tilde{\mu}_{\sigma} &= \mu(1+n_{\bar{\sigma}}) + \sigma h + (V-\frac{J}{2})\langle n_{i\bar{\sigma}}\rangle \nonumber\\
& \ \ \ \ \ \ \ \ \ \ \ \ \ \  + 2 t (\langle c^{\dagger}_{i+\hat{e}_1,\bar{\sigma}} c_{i,\bar{\sigma}}\rangle+\text{c.c.}) 
\end{align}
Once the variational parameters $\tilde{t}_\sigma$ and $\tilde{\mu}_\sigma$ are self-consistently solved, and the solution with the lowest mean-field energy is determined, one can obtain the simple dispersion of the renormalized bands:
\begin{align}
    &\epsilon_\sigma(\vec{k}) + \tilde{\mu}_\sigma \nonumber\\  =&  2|\tilde{t}_\sigma|\left[\cos(k_x+\theta_\sigma) + 2\cos\left(\frac{\sqrt{3}}{2}k_y\right)\cos\left(\frac{k_x}{2}-\theta_\sigma\right)\right]
\end{align}
where we have decomposed $\tilde{t}_\sigma = |\tilde{t}_\sigma|\mathrm{e}^{\mathrm{i} \theta_\sigma}$. We see that the complex phase of the hopping element plays the role of energetically distinguishes the two valleys while keeping the positions of the band minima at $K$ points.
\section{The Eigenvalues of $M$}
\begin{align}
    \lambda_0 &= \frac{21}{2}V-\frac{15}{4}J+6|t|+ \frac{1}{2\rho_0} \\
    \lambda_\text{s} &= -\frac{3}{2}V+\frac{15}{4}J-6|t|+ \frac{1}{2\rho_0} \\
    \lambda_\text{v} &= -\frac{9}{2}V-\frac{9}{4}J+ \frac{1}{2\rho_0} \\
    \lambda_\text{sv} &= -\frac{9}{2}V+\frac{9}{4}J+ \frac{1}{2\rho_0}
\end{align}

\section{Magnetic response in the other subregions of valley polarized phase}

Here we briefly comment on the magnetic response of other subregions of the V phase. When $0<\tilde{\lambda}<|\lambda_{\text{V}}|<\lambda_\text{SVL}$, the responses in the first and the third stages remain unchanged from the above case. However, the intermediate stage in Fig.~2 of the main text collapses to a sudden change (the lines in the corresponding figure would be vertical) at a critical magnetic field
\begin{align}
    h_c = \frac{n}{2}\left[\lambda_\text{S}+\lambda_\text{SVL}-\sqrt{(\lambda_\text{S}+\lambda_\text{SVL})(\lambda_\text{V}+\lambda_\text{SVL})}\right]
\end{align} 
This magnetic response can be viewed as a special case of the case we have focused on in the main text. When $\lambda_{\text{S}}<0$ or $\lambda_{\text{SVL}}<|\lambda_\text{V}|$, the system will either be quickly spin polarized in a two-stage process, or eventually become flavor polarized at high field $H$. Therefore, although these two cases can have rather complicated and interesting magnetic response, they are irrelevant to the physics we are discussing in this paper.

\end{document}